\documentclass[manuscript]{aastex}
\newcommand{\be}{\begin{equation}}
\newcommand{\ee}{\end{equation}}
\newcommand{\bse}{\begin{subequations}}
\newcommand{\ese}{\end{subequations}}
\newcommand{\bary}{\begin{eqnarray}}
\newcommand{\eary}{\end{eqnarray}}

\shorttitle{SSC EMISSION FROM REVERSE SHOCK}
\shortauthors{Fraija, Gonz\'{a}lez, \& Lee}

\begin{document}
\title{SSC EMISSION  AS THE ORIGIN OF THE  GAMMA RAY AFTERGLOW OBSERVED IN GRB 980923}
\author{N. Fraija, M. M. Gonz\'alez and  W. H. Lee}
\affil{Instituto de Astronom\'ia, Universidad Nacional Aut\'{o}noma de M\'{e}xico, Apdo. Postal 70-264, Cd. Universitaria, M\'{e}xico DF 04510}
\email{nifraija@astro.unam.mx, mgonzalez@astro.unam.mx,wlee@astro.unam.mx}


\begin{abstract}

GRB 980923 was one of the brightest bursts observed by the Burst and Transient Source Experiment (BATSE). Previous studies have detected two distinct components in addition
to the main prompt episode, which is well described by a Band function. The first of these is a tail with a duration of $\simeq 400$s, while the second is a high-energy component lasting $\simeq 2$~s.
We summarize the observations, and argue for a unified model in which the tail can be understood as the early $\gamma$-ray afterglow from forward shock synchrotron emission, while the high-energy component arises from synchrotron self-Comtpon (SSC) from the reverse shock. Consistency between the main assumption of thick shell emission and agreement between the observed and computed values for fluxes, break energies, starting times and spectral indices   leads to a requirement that the ejecta be highly magnetized. 
\end{abstract}


\section{Introduction}

The most successful theory in terms of explaining GRBs and their afterglows is the fireball model \citep[see][for recent reviews]{mes06, zha04}.  This model predicts  an  expanding  ultrarelativistic shell that moves into the external surrounding medium.  The collision of the expanding shell  with another shell (internal shocks) or the interstellar medium (external shocks) gives rise to radiation emission through the synchrotron and SSC processes.  In addition,  when  the expanding relativistic shell encounters the external medium two shocks are involved: an outgoing, or forward, shock \citep{ree94,pac93} and another one that propagates back into the ejecta, the reverse shock  \citep{mes94, mes97a}.\\ 
According to the standard relativistic fireball model, the forward shock accelerates electrons (through first order Fermi mechanism or electric fields associated with the Weibel instability) up to relativistic energies and generates magnetic fields \citep{med99}. The afterglow emission is more likely to be synchrotron \citep{sar98}; however inverse Compton scattering may affect the electron radiative cooling. The progressive, power-law deceleration of the forward shock leads to a continuous softening of the afterglow spectrum \citep{pan07}.   This has been observed in the spectra and light curves of  GRB 970228\citep{sar98}, GRB 970508\citep{pan98, sar98}, GRB 980923 \citep{gib99}  and among a dozen  other afterglows \citep{zeh06}.\\
On the other hand, the  reverse shock is predicted to produce a strong optical flash \citep{mes97a, sar99a,sar99b}. When it crosses the shell, the reverse shock heats it up and accelerates electrons, but it operates only once.  Thus, unlike the forward shock emission that continues later at lower energy, the reverse shock emits a single burst. After the peak of the reverse shock no new electrons are injected and the shell material cools adiabatically \citep{ven09}. However, this picture can be modified allowing a long-lived reverse shock if the central engine emits slowly moving material in which the reverse shock could propagate and survive for hours to days \citep{gen07,uhm07}.  Now, although the contribution of the reverse shock  synchrotron emission to the X-ray band is small, electrons in the reverse shock region can upscatter the synchrotron photons (SSC process) up to the X-ray band or even higher energies \citep{mes93, wan01a, wan01b,wan05,gra03, kob07a,zha06}.\\

The soft tail component seen in GRB980923 has been described reasonably well by synchrotron emission from a decelerating relativistic shell that collides with an external medium (forward shock model), and also described as such in more than a dozen independent events \citep[see, e.g.][]{che99, rho99, sar98, pan07, gib99,sha05}. The high energy component, on the other hand, has been explored through inverse Compton emission models \citep{gra03,pee04} and SSC emission from the reverse shock \citep{wan05,kob07a,wan01b}. In all cases a full description of the high energy component  features has not been achieved.

In this paper we present a unified description of the tail and the high energy component in GRB980923 through a forward and reverse shock, accounting for their energy, spectral indices, fluxes and duration. In order to do so consistently we find that a magnetized outflow is required. 

\section{GRB 980923}

\begin{figure}
\epsscale{.80}
\plotone{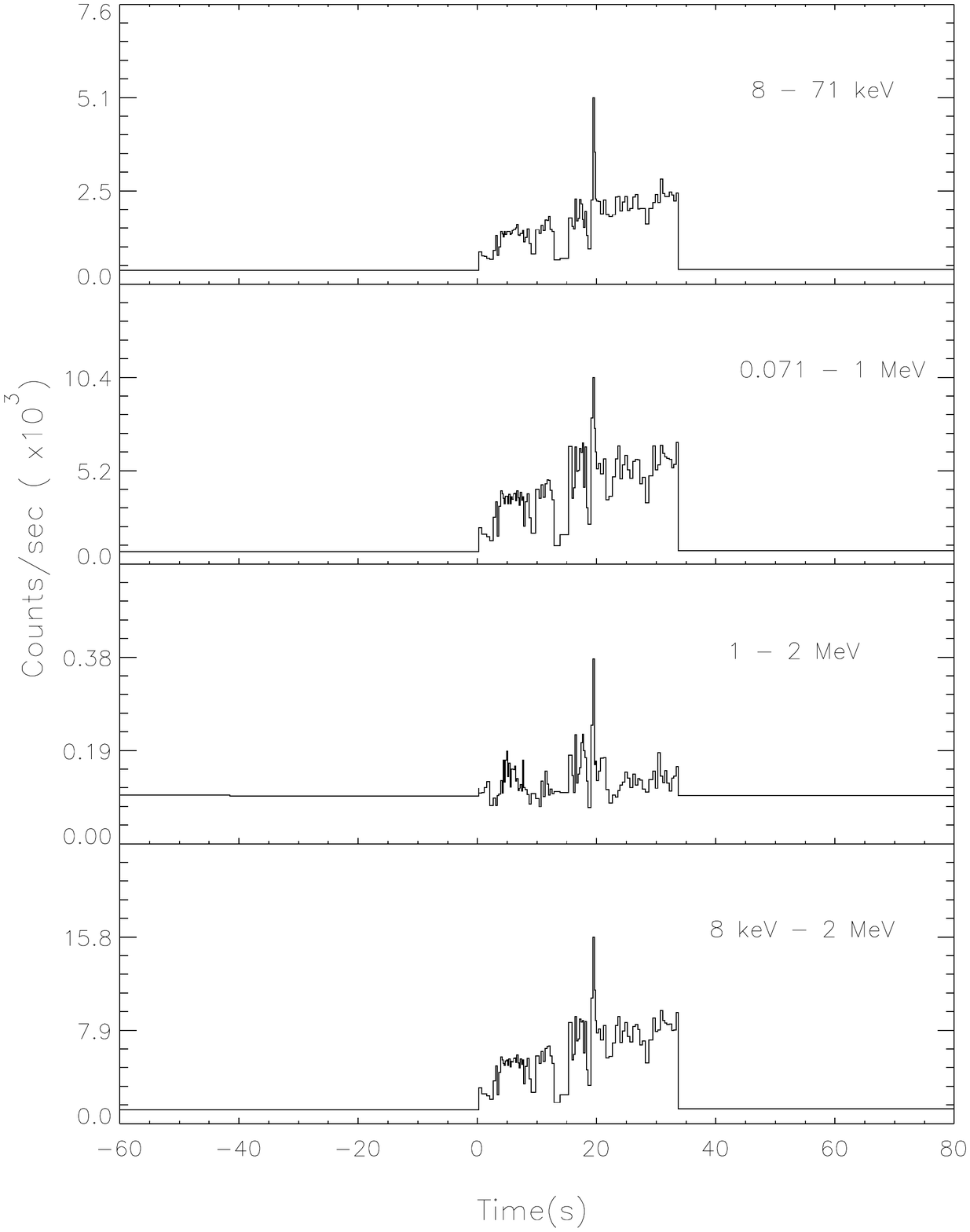}
\caption{The BATSE SD7 count rates for GRB980923 are plotted in four energy ranges. A peak at 20~s is apparent, and stands out particularly at the low and high energy ranges.}
\label{fig2}
\end{figure}

GRB 980923 was observed by  the Burst and Transient Source Experiment (BATSE) on 1998 September 23 at 20:10:52 UT for 32.02 s. It was  localized  to $234^0$ with respect to the pointing-axis direction of CGRO. 
In agreement with the lightcurve given by \citet{gon09,gon11} (Figure~\ref{fig2}), the event consists of three components.  The first is related to the typical prompt emission, the second one  to a smooth tail which lasts $\sim\,400$s  and the last to a hard component extending up to $\sim 150\, MeV$.  The smooth tail was well described by  \citet{gib99} as the evolution of a synchrotron cooling break in the  slow-cooling regime starting at $t_0=32$ seconds. This conclusion was  based on two  different aspects. First,  the time-dependent break energy was modeled with a power law of the form $E_{0}(t-t_{0})^{\delta}$ and for $t_0=32.109\,$ seconds,  they obtained $\delta=-0.52\pm 0.12$.   Second,  in agreement with the computed value from the spectral index  $p=2.4\pm 0.11$ and  the value of $t_0$, they obtained the relationship between the temporal and spectral indices  and observed that the index values  were indeed closer to the slow cooling regime.    Thus, \citet{gib99}  identified the evolution of the spectrum in the tail of the burst as the evolution of a synchrotron cooling break in the slow-cooling regime, implying also that the transition from fast to slow cooling  could take place on short  timescales, comparable to the duration of the burst.  On the other hand, \citet{gon09,gon11} described the high energy component peaking at $t=20s$ as a power law with spectral index,  $\gamma$ of $1.439\pm 0.0687$ and flux $F=49.2 \pm 3.8$~erg~cm$^2$~s$^{-1}$ and also  pointed out  that the tail could begin before or at least  about 14s after the burst trigger.  We assume  
 \citet{gon09,gon11}  results to develop a unified model where both components are related.  

\section{Dynamics of the forward and reverse shock} 

In this section, we first extend the work  done by \citet{gib99} on the smooth tail, and then proceed to compute the energy range for SSC emission from a thick shell of the reverse shock fireball to account for the hard component. We point out that the tail could have begun before or at least  about 14s after the burst trigger, suggesting that before this time there already existed  forward and reverse shocks, and find that the characteristics of the observed hard emission noted by \citet{gon09,gon11} can be accounted for consistently  by a scenario where the reverse shock becomes relativistic during its propagation, with magnetization parameter close to unity.
The subscripts $\rm f$ and $\rm r$ refer throughout to the forward and reverse shock, respectively. 

\subsection{Choice of $t_0$}
In general $t=0$ is defined by the trigger time of the burst.  Because of the connection between the very early afterglow (smooth tail) and this prompt emission, we plotted  the power index $\delta$ as function of $t_0$ ( Fig. \ref{fig1}) for the data and the power law of the form $E_{0}(t-t_{0})^{\delta}$ \citep{gib99}.   It is clear that  $\delta$ goes from fast cooling regime to slow cooling regime depending on the choice of $t_0$. Also, we observe as a particular case  that for $t_0=32$s,  $\delta\sim 0.52$ is obtained (i.e. slow cooling).  However, \citet{gon09,gon11} points out  that the tail could begin before or at least  about 14s after the burst trigger.  From Fig. \ref{fig3} we observe that the data  used  to define $t_0$ and $\delta$ are not very restrictive, so values of $t_0=14$s and $\delta\sim 0.9$ are consistent the data too. 
Hence,  choosing $t_0=14$s  gives a $\delta$ value in the transition from fast to slow cooling that can be explain if the transition time is sufficient short so that the spectra would be never observe in the fast cooling regime (this is the case we calculate later). we can assume that about this time the GRB ejecta  collides with the ISM generating  forward and reverse shocks.

\subsection{Smooth tail from Synchrotron radiation forward shock}

For the forward shock, we assume that electrons are accelerated in the shock to a power law distribution of Lorentz factor $\gamma_e$ with a minimum Lorentz factor $\gamma_m$: $N(\gamma_e)d\gamma_e\propto \gamma_e^{-p}d\gamma_e$,
$\gamma_e\geq\gamma_m$ and  that constant fractions $\epsilon_{e,f}$ and $\epsilon_{B,f}$ of the
shock energy go into the electrons and the magnetic field, respectively. Then
\begin{eqnarray}
\gamma_{m,f}&=&\epsilon_{e,f}\biggl(\frac {p-2} {p-1}\biggr) \frac {m_p} {m_e}\gamma_f\cr
&=&524.6 \,\epsilon_{e,f}\,\gamma_{\rm f}
\end{eqnarray}
where we have used the value of $p=2.4\pm 0.11$ as obtained by \citet{gib99}.  Adopting the notation of \citep{sar98} and ignoring self-absorption,  the observed spectral flux in the fast-cooling regime is given by 

\begin{equation}
\label{fcsyn}
F_\nu=
F_{\nu,\rm max}
\cases{ 
(\nu/\nu_{\rm c,f})^{1/3},& $\nu_{\rm c,f}>\nu$ \cr
 (\nu/\nu_{\rm c,f})^{-1/2}, & $\nu_{\rm c,f}<\nu<\nu_{\rm m,f}$, \cr
(\nu_{\rm m,f}/\nu_{\rm c,f})^{-1/2} 
( \nu/\nu_{\rm m,f})^{-p/2}, & $\nu>\nu_{\rm m,f}$. \cr
}
\end{equation}

\noindent Similarly, the flux in the slow-cooling regime can be written
as

\begin{equation}
\label{scsyn}
F_\nu=
F_{\nu,\rm max}
\cases{ 
(\nu/\nu_{\rm m,f})^{1/3},& $\nu_{\rm m,f}>\nu$ \cr
(\nu/\nu_{\rm m,f})^{-(p-1)/2}, & $\nu_{\rm m,f}<\nu<\nu_{\rm c,f}$, \cr
(\nu_{\rm c,f}/\nu_{\rm m,f})^{-(p-1)/2} 
(\nu/\nu_{\rm c,f})^{-p/2}, & $\nu>\nu_{\rm c,f}$. \cr
}
\end{equation}

Using the typical parameters given by \citet{bjo01},  we compute the typical and cooling frequencies of the forward shock synchrotron emission  \citep{sar98} which are given by,
\bary\label{synforw}
\nu_{\rm m,f}&\sim& 1.9 \times 10^{19} \biggl(\frac{1+z}{2}\biggr)^{1/2}\,\biggl(\frac{\epsilon_{e,f}}{0.95}\biggr)^2\,\epsilon^{1/2}_{B,f,-5}\,E^{1/2}_{54}\,t^{-3/2}_{1}\,{\rm \ Hz}\cr
\nu_{\rm c,f}&\sim& 3.0 \times 10^{19}\biggl(\frac{1+z}{2}\biggr)^{-1/2}\,\biggl(\frac{1+x_f}{2.5}\biggr)^{-2}\,\epsilon^{-3/2}_{B,f,-5}\,n^{-1}_{f,0}\,E^{-1/2}_{54}\,t^{-1/2}_{1}\, {\rm \ Hz}\cr
F_{\rm max,f}&\sim& 2.2\times 10 \biggl(\frac{1+z}{2}\biggr)\,\epsilon^{1/2}_{B,f,-5}\,n^{1/2}_{f,0}\,D^{-2}_{28}\,E_{54}\,{\rm \,\mu Jy}\cr
t_{\rm tr,f}&\sim& 8.7 \biggl(\frac{1+z}{2}\biggr)\, \biggl(\frac{\epsilon_{e,f}}{0.95}\biggr)^2  \,\epsilon^{2}_{B,f,-5}\,n_{f,0}\,E_{54}\,{\rm s}
\eary

\noindent where the convention $Q_x=Q/10^x$ has been adopted in cgs units throughout this document unless otherwise specified. $t_{\rm tr,f}$ is the transition time, when the spectrum changes from fast cooling to slow cooling,  $D$ is the luminosity distance, $n_{f}$ is the ISM density, $t$ is the time of the evolution of the tail, $E$ is the energy, and the term $(1+x_f)$ was introduced because a once-scattered synchrotron photon generally has energy larger than the electron mass in the rest frame of the second-scattering electrons.  Multiple scattering of synchrotron photons can be ignored. $x_f$ is given by \citep{sar01} as:

\begin{equation}
x_f = \left\{ \begin{array} {ll} 
\frac{\eta \epsilon_{e,f}}{\epsilon_{B,f}}, & \mathrm{if \quad}
\frac{\eta \epsilon_{e,f}}{\epsilon_{B,f}} \ll 1, \\ 
\left(\frac{\eta \epsilon_{e,f}}{\epsilon_{B,f}}\right)^{1/2}, & \mathrm{if \quad}
\frac{\eta \epsilon_{e,f}}{\epsilon_{B,f}} \gg 1. 
\end{array} \right.
\end{equation}
where $\eta=(\gamma_{\rm c,f}/\gamma_{\rm m,f})^{2-p}$ for slow cooling and $\eta=1$ for fast cooling.\\
From eq. \ref{synforw}, we observe directly that $\nu_{\rm m,f}  \leq \nu_{\rm c,f}$, the break energy $E_{\rm c,f}\sim 124.1\,$keV  is  consistent with the values given by  \citep{gib99} and  $t_{\rm tr,f}\sim 8.7$s, implying also that the transition from fast to slow cooling  could take place on  very short  timescales, comparable to the duration of the burst, as expected

\subsection{ X ray flare from thick shell  Reverse shock}
For the reverse shock, it is possible to obtain a simple analytic solution in two limiting cases, thin  and thick  shell,  \citep{sar95} by using a critical Lorentz factor $\Gamma_c$, 

\bary
\Gamma_c&=&\biggl(\frac{3E}{4\pi n_rm_p c^5 T^3}\biggr)^{1/8}\biggl(\frac{1+z}{2}\biggr)^{3/8}\cr 
&=&255.2\,\biggl(\frac{1+z}{2}\biggr)^{3/8}\,n^{-1/8}_{r,1}\,E^{1/8}_{54}\,\biggl(\frac{T_{90}}{32s}\biggr)^{-3/8}
\eary
where $T_{90}$ is the time of the GRB, which is much larger that the peak time of the reverse shock emission, and $n_r$ is the thick shell density. We consider the thick shell case in which the reverse shock becomes relativistic during the propagation and the shell is significantly decelerated by the reverse shock.  Hence, the Lorentz factor at the shock crossing time $t_c$ is given  by $\gamma_d\sim \Gamma_c$ \citep{kob07a,kob07b}, and for  $\sigma \sim 1$, where $\sigma=L_{pf}/L_{kn}= B_r^2/4\pi n_r m_p c^2 \Gamma_r^2$ is the magnetization parameter, defined as the ratio of Poynting flux to matter energy flux, the crossing time $t_c$ is much shorter  than $T_{90}$, $t_c\sim T_{90}/6$,  \citep{fan04b, fan08, zha05, dre02}.   Now, if the constant fractions, $\epsilon_{e,r}$ and $\epsilon_{B,r}$ of the reverse shock energy go into the electrons and magnetic fields, respectively, we have
\bary
\gamma_{\rm m,r}&=&\epsilon_{e,r}\biggl(\frac {p-2} {p-1}\biggr) \frac {m_p} {m_e}\frac{\gamma_r}{\Gamma_c}\cr
&=& 1233.5\,\biggl(\frac{1+z}{2}\biggr)^{-3/8}\,\biggl(\frac{\epsilon_{e,r}}{0.6}\biggr)\,\gamma_{\rm r,3}\,n^{1/8}_{r,1}\,E^{-1/8}_{54}\,\biggl(\frac{T_{90}}{32s}\biggr)^{3/8}
\eary
where $\gamma_r$ is the Lorentz factor of the thick shell. The spectral characteristics of the forward and reverse shock synchrotron emission are related \citep{zha03,kob07a,fan05,fan04a,jin07,sha05} by,

\bary\label{conec}
\nu_{\rm m,r}&\sim&\,\mathcal{R}^2_e\,\mathcal{R}^{-1/2}_B\,\mathcal{R}^{-2}_M\,\nu_{m,f}\cr
\nu_{\rm c,r}&\sim&\,\mathcal{R}^{3/2}_B\,\mathcal{R}^{-2}_x\,\nu_{c,f}\cr
F_{\rm max,r}&\sim&\,\mathcal{R}^{-1/2}_B\,\mathcal{R}_M\,F_{max,f}
\eary
where $\mathcal{R}_B=\epsilon_{B,f}/\epsilon_{B,r}\,$, $\mathcal{R}_e=\epsilon_{e,r}/\epsilon_{e,f}\,$,  $\mathcal{R}_x=(1+x_f)/(1+x_r+x_r^2)$ and $\mathcal{R}_M=\Gamma^2_c/\gamma$.  The previous relations tell us that including the re-scaling there is a unified description between both shocks (forward and reverse), and the distinction between forward and reverse magnetic fields considers that in some central engine models \citep{uso92,mes97b,whe00}  the fireball wind may be endowed with ``primordial" magnetic fields. Also as the cooling Lorentz factor must be corrected, then  $\mathcal{R}_x$ is introduced as a correction factor for the IC cooling, where $x_r$ is obtained by \citep{kob07a} as,

\begin{equation}
x_r = \left\{ \begin{array} {ll} 
\frac{\eta \epsilon_{e,r}}{\epsilon_{B,r}}, & \mathrm{if \quad}
\frac{\eta \epsilon_{e,r}}{\epsilon_{B,r}} \ll 1, \\ 
\left(\frac{\eta \epsilon_{e,r}}{\epsilon_{B,r}}\right)^{1/3}, & \mathrm{if \quad}
\frac{\eta \epsilon_{e,r}}{\epsilon_{B,r}} \gg 1. 
\end{array} \right.
\end{equation}

For fast cooling , we take $\eta=1$, and hence with the standard values for $\epsilon_{B,r}$ and $\epsilon_{e,r}$, $\eta\epsilon_{B,r}/\epsilon_{e,r}\sim 4.8$. Using equations (\ref{synforw}) and (\ref{conec}), the typical and cooling frequencies of the reverse shock synchrotron emission are
\bary\label{synrev}
\nu_{\rm m,r}&\sim& 3.4\times 10^{16}\biggl(\frac{1+z}{2}\biggr)^{-1}\,\biggl(\frac{\epsilon_{e,r}}{0.6}\biggr)^{2}\,\biggl(\frac{\epsilon_{B,r}}{0.125}\biggr)^{1/2}\,\gamma^{2}_{r,3}\,n^{1/2}_{r,1}\, {\rm \ Hz}, \cr
\nu_{\rm c,r}&\sim& 1.5\times 10^{11} \biggl(\frac{1+z}{2}\biggr)^{3/2}\,\biggl(\frac{1+x_r+x_r^2}{6}\biggr)^{-2}\,\biggl(\frac{\epsilon_{B,r}}{0.125}\biggr)^{-7/2}\,n^{-3}_{r,1}\,E^{-1/2}_{54}\,\gamma^{-6}_{r,3}\,\biggl(\frac{T_{90}}{32s}\biggr)^{5/2}\, {\rm \ Hz}, \cr
F_{\rm max,r}&\sim& 5.02\times 10^2 \biggl(\frac{1+z}{2}\biggr)^{7/4}\,\biggl(\frac{\epsilon_{B,r}}{0.125}\biggr)^{1/2} \,n^{1/4}_{r,1}\,D^{-2}_{28}\,E^{5/4}_{54}\,\gamma^{-1}_{r,3}\,\biggl(\frac{T_{90}}{32s}\biggr)^{-3/4}\,{\rm \,Jy}.
\eary

From equation~\ref{synrev} we see that $\nu_{\rm m,r}$ and $\nu_{\rm c,r}$  correspond to optical and IR frequencies, respectively, and that  $\nu_{\rm m,r}$, which characterizes the frequency band, does not depend on $x_r$. However  these energies  were not recorded. Instead, as higher energy photons were observed we compute the upscattering emission of the synchrotron radiations (equations~\ref{fcsyn} and \ref{scsyn}) ($r\to$f) by relativistic electrons (fast cooling and slow cooling \citep{sar01}). So the SSC spectrum in the fast cooling regime is, 

\begin{equation}\label{nuFnu_SSC_FC}
\nu F^{\rm SSC}_{\nu}
=
(\nu F_\nu)^{\rm SSC}_{\rm max}
\cases{
(\nu^{\rm IC}_c / \nu^{\rm IC}_m)^{1/2} (\nu / \nu^{\rm IC}_c)^{4/3}			 		&  $\nu<\nu^{\rm IC}_c$\cr
(\nu/\nu^{\rm IC}_m)^{1/2}                                                                                                        	&  $\nu^{\rm IC}_c<\nu<\nu^{\rm IC}_m$ \cr
(\nu/\nu^{\rm IC}_m)^{(2-p)/2}                                                                                                          &  $\nu>\nu^{\rm IC}_m $ \cr
}
\end{equation}
and in the slow cooling regime we find
\begin{equation}\label{nuFnu_SSC_SC}
\nu F^{\rm SSC}_{\nu} 
=
(\nu F_\nu)^{\rm IC}_{\rm max}
\cases{
(\nu^{\rm IC}_m / \nu^{\rm IC}_c)^{(3-p)/2} (\nu / \nu^{\rm IC}_m)^{4/3} 				& $\nu<\nu^{\rm IC}_m$ \cr 
(\nu/\nu^{\rm IC}_c)^{(3-p)/2} 												& $\nu^{\rm IC}_m<\nu<\nu^{\rm IC}_c$  \cr
(\nu/\nu^{\rm IC}_c)^{(2-p)/2} 												& $\nu>\nu^{\rm IC}_c$ \cr
}
\end{equation}

\noindent $\nu^{(\rm IC)}$, $\nu^{(IC)}_{c}$ and $F^{(IC)}_{max}$ are given by

\bary\label{ic}
\nu^{(\rm IC)}_{m}\sim\gamma^2_{m},\nu_{m,r}\,;\hspace{2cm}\nu^{(IC)}_{c}\sim\gamma^2_c\,\nu_{c,r}\,;\hspace{2cm}F^{(IC)}_{max}\sim\,k\tau\,F_{\rm max,r};
\eary
where $k=4(p-1)/(p-2)$ and $\tau=\frac{\sigma_T N_e}{4\pi R_d}=\frac{c}{3} \biggl(\frac{1+z}{2}\biggr)^{-1}\,\sigma_T\,n\,\Gamma_c^4\, \gamma^{-1}\,$ is the optical depth of the shell.  In agreement with equations~(\ref{synrev}) and (\ref{ic}), in the Self-Synchrotron Compton we have,

\bary\label{ssc}
\nu^{(IC)}_{\rm m}&\sim& 1.034\times 10^{23} \biggl(\frac{1+z}{2}\biggr)^{-7/4}\,\biggl(\frac{\epsilon_{e,r}}{0.6}\biggr)^{4}\,\biggl(\frac{\epsilon_{B,r}}{0.125}\biggr)^{1/2}\,\gamma^{4}_{r,3}\,n^{3/4}_{r,1}\,E^{-1/4}_{54}\,\biggl(\frac{T_{90}}{32s}\biggr)^{3/4}\, {\rm \ Hz},\cr
\nu^{(IC)}_{\rm c}&\sim& 1.1\times 10^{10} \biggl(\frac{1+z}{2}\biggr)^{3/2}\,\biggl(\frac{1+x+x^2}{6}\biggr)^{-4}\,\biggl(\frac{\epsilon_{B,r}}{0.125}\biggr)^{-7/2}\,n^{-3}_{r,1}\,E^{-1/2}_{54}\,\gamma^{-6}_{r,3}\,\biggl(\frac{T_{90}}{32s}\biggr)^{-5/2}\, {\rm \ Hz},\cr
F^{(IC)}_{\rm max}&\sim& 4.7 \times 10^{-2} \biggl(\frac{1+z}{2}\biggr)^{9/4}\,\biggl(\frac{\epsilon_{B,r}}{0.125}\biggr)^{1/2}\,n^{3/4}_{r,1}\,D^{-2}_{28}\,E^{7/4}_{54}\,\gamma^{-2}_{r,3}\,\biggl(\frac{T_{90}}{32s}\biggr)^{-5/4}\,{\rm \,Jy}.
\eary

From equation~\ref{ssc} we observe that the break energies and $(\nu F)_{\rm max}=21.2\times 10^{-6}$~erg~cm$^{-2}$~s$^{-1}$ are within the range pointed out by \citet{gon09,gon11}.

\section{Discusion and Conclusions}
\begin{figure}
\epsscale{.80}
\plotone{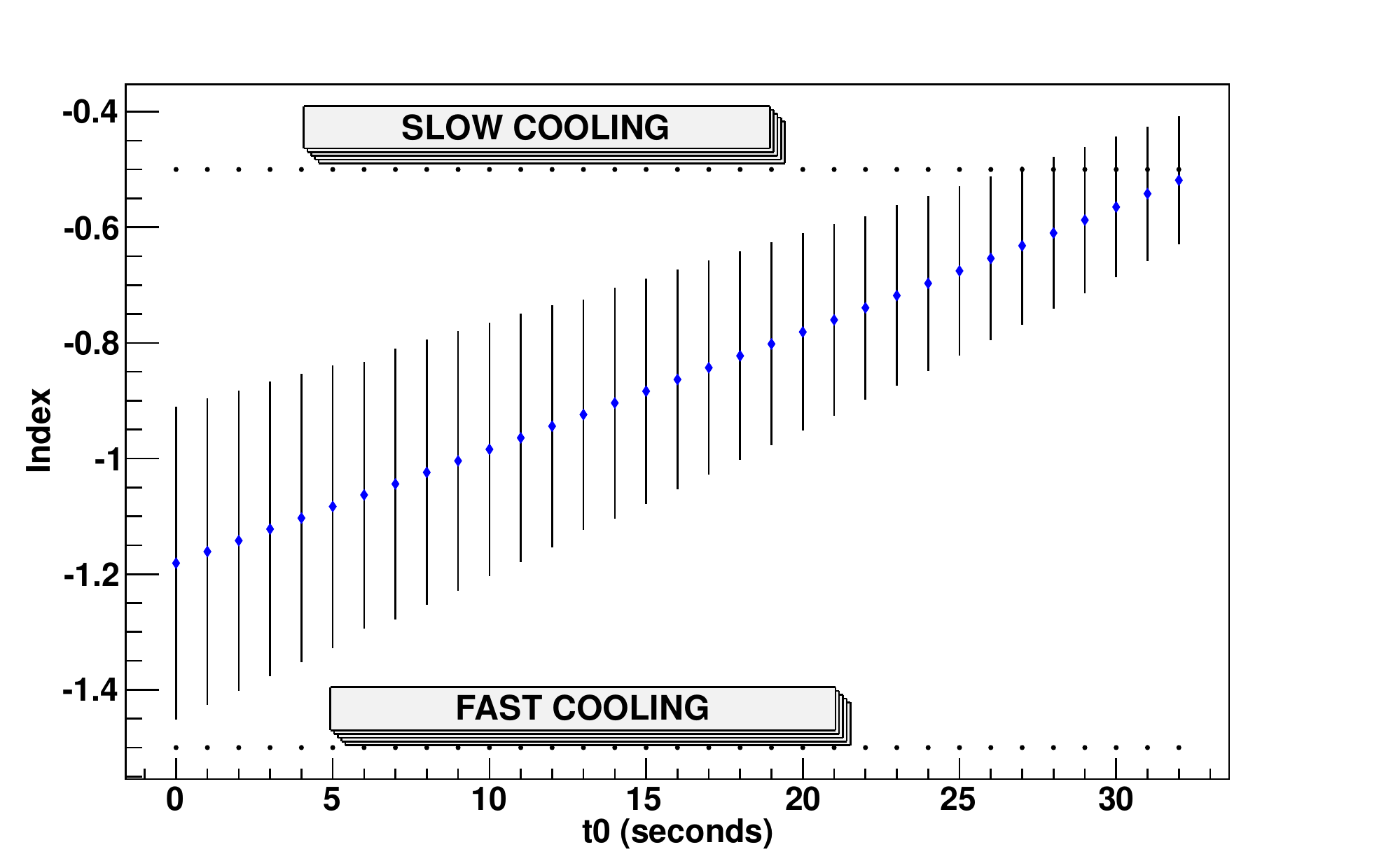}
\caption{The power-law decay index $\delta$ is plotted as a function of $t_0$ for a power law of the form $E_0\,(t - t_0)^\delta$, where $t_0$ is the time at which the tail begins. The error bars are obtained for the different fits, as shown in figure \ref{fig3}.   The index is sensitive to the value $t_0$, and thus the choice of $t_0$ is crucial and  illustrates the effect on the light-curves and therefore on the apparent evolution of the spectrum. The expected values for the slow and fast cooling regimes are indicated.}
\label{fig1}
\end{figure}

\begin{figure}
\epsscale{.80}
\plotone{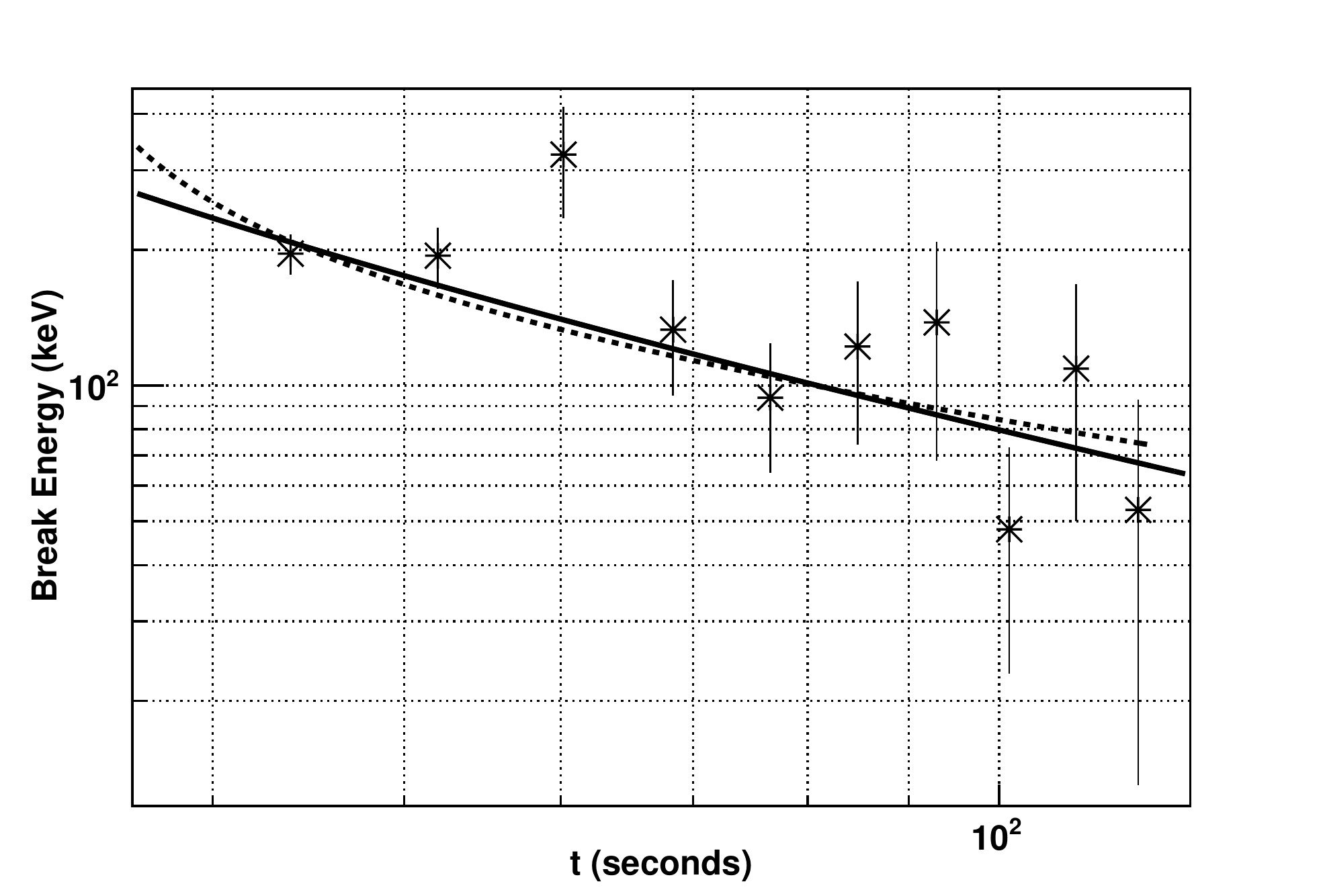}
\caption{Break energy as a function of time $t$ on logarithmic axes. The continuous line represents the fitted power law slope 0.5186$\pm$0.1113  for $t_0$=32 s \citep{gib99}, while the dashed line is the fitted power law slope 0.9039$\pm$0.1999  for $t_0$=14 s. $t=0$ corresponds to the burst trigger. }
\label{fig3}
\end{figure}

As shown in Figure~\ref{fig1}, the power law decay index is sensitive to the chosen value of $t_0$, so depending on our choice,  we will be apparently closer to a fast or slow cooling regime. Choosing $t_0\leq14$s \citep{gon09,gon11,sac11a}, and assuming that from this time until $t\sim32$s the synchrotron emission was eclipsed by the prompt phase, we have calculated  the transition time from fast to slow cooling as $\sim 8.7\,$s. Thus,  18s later, at $t=32$s  the synchrotron process generated by the forward shock was in the slow-cooling regime and in the energy range corresponding to that reported by \citet{gib99}. Also, we suggest that  the diminishing flux at $\simeq14$~s may be due to pair production  ($\gamma\gamma\to e^+e^-$) between prompt emission  and  forward shock photons (equation \ref{synforw})  at the beginning of the afterglow.   We have calculated the energy of the forward shock photons as  $1.31\,$MeV and the time for this energy  to decrease under the pair production umbral as $1.5 s$  which is consistent with duration of the diminishing flux. \\

In the reverse shock, the synchrotron process emitted photons with $\nu_{\rm c,r}\sim 1.0\times 10^{10}\,$Hz and $\nu_{\rm m,r}\sim4.6\times 10^{16}\,$Hz, which were not recorded but were  upscattered by electrons  up to break energies $E^{IC}_{\rm c}\sim 4.2\times 10^{-5} \,$eV, $E^{IC}_{\rm m}\sim\,427.9\,$MeV with a  $(\nu F)_{\rm max}=21.2\times 10^{-6}$~erg~cm$^{-2}$~s$^{-1}$, which were pointed out by \citet{gon09,gon11}  and  \citet{sac11a}.   Now,  in accordance with the observed value for the spectral index $\gamma\sim 1.44\pm 0.07$,  $\nu^{(ic)}_{m,r}>\nu^{(ic)}_{c,r}$   we conclude the  SSC spectrum corresponds to fast-cooling regime, very similar to GRB 941017 \citep{gon03,gra03}. For our case (thick case), the flare occurs during the prompt gamma-ray phase.\\ 
From the value of  $\mathcal{R}_B$, we obtained that the forward and reverse magnetic fields are related by  $B_f=0.9\times 10^{-3}B_r$.  The previous result indicates  that there is a stronger magnetic field in the reverse-shock region than in the forward shock region, which may suggest that the obtained  results are given when the ejecta is highly magnetized, as  in the interpretation of the early afterglow of GRB 990123 and GRB 021211 provided  by \citet{zha03}.\\
Finally, because the Large Area Telescope (LAT) covers the energy range from about 20 MeV to more than 300 GeV,  we hope to detect other hard components in GRBs and so further constrain this model.

The current model accounts for the main characteristics of the burst: energies, spectral indices, fluxes, duration of the main components in a unified manner. The main requirements are that the ejecta be magnetized, leading to the formation of a reverse shock. The model has eight free parameters (equipartition magnetic field, equipartition electron energy, Lorentz factor, and densities all of the in the reverse and forward shocks), with standard values.  The main difference between our model and previous models \citep{wan05,kob07a,wan01b}  are the assumption  of different equipartition values in the forward and reverse shock which leads the magnetization of the jet.

This burst has similar characteristics to GRB 090926A~\citep{ack11}, but the high energy extended emission component may require SSC from the forward shock \citep{sac11b}.

\acknowledgments

We thank Enrico Ram\'{\i}rez-Ruiz and  Charles Dermer for useful discussions. This work is partially supported by UNAM-DGAPA PAPIIT IN105211 (MG) and CONACyT 103520 (NF) and 83254 (WL).

\clearpage

\end{document}